\documentstyle[prl,aps]{revtex}
\input{epsf}
\draft
\begin{document}
\twocolumn[\hsize\textwidth\columnwidth\hsize\csname@twocolumnfalse\endcsname

\title{Angular Dependence of $C$-Axis Magnetoresistance in
${\rm Bi_2Sr_2CaCu_2O_{8+
\delta}}$ Single Crystals With Columnar Defects}
\author{N. Morozov$^{a}$, L.~N. Bulaevskii$^{a}$, M.~P. Maley$^{a}$,
and J.~Y. Coulter$^{a}$\\
A.~E. Koshelev $^{b}$ and T.~-W. Li$^{b}$}
\address{$^{a}$Los Alamos National Laboratory,
MST-STC, MS-K763, Los Alamos, NM 87545\\
$^b$  Argonne National Laboratory, MSD, Argonne, IL 60439}
\date{\today}
\maketitle

\begin{abstract}

We measured the angular dependence of the $c$-axis
magnetoresistance $\rho_{c}({\bf B})$ of ${\rm
Bi_2Sr_2CaCu_2O_{8+\delta}}$ irradiated with heavy ions.  At
temperatures near 68~K the scaling of $\rho_{c}({\bf B})$ with the
$c$-axis magnetic field component $B_{\bot}$ is broken and the in-plane field,
$B_{\parallel}$, affects $\rho_{c}$.  At this temperature, at a
specific field $B_{cr}\approx B_{\Phi}/2$, magnetoresistance becomes
independent of field
orientation.  This crossing point allows us to estimate the
correlation length $L$ of pancake positions along the $c$ axis.
We find $L/s\sim 100$ at $B=B_{cr}$, where $s$ is the interlayer spacing.
This provides evidence of strong enhancement of pancake alignment in the vortex
liquid in crystals with columnar defects.

\end{abstract}


]
\narrowtext

The properties of the vortex liquid in high-temperature
superconductors, HTS, in the presence of strong disorder is one of the
most challenging problems in the physics of the vortex state.  It is
well established now that in pristine
${\rm Bi_2Sr_2CaCu_2O_{8+\delta }}$, Bi-2212, crystals vortices
form a weakly $c$-axis correlated pancake liquid.  Thermodynamics and
intralayer dynamics of
this liquid are defined mostly by pancake concentration, i.e., by the
$c$-axis magnetic field $B_{\bot}$ \cite{kes-vl}.  This behavior is
however, strongly affected by the presence of correlated disorder.
The most effective pinning centers are produced by heavy ion
irradiation.  Irradiation of the HTS by energetic ions produces
amorphous tracks, where the superconductivity is suppressed.  Such
columnar defects, CDs, with radii comparable with the superconducting
coherence length, are ideal for pinning vortex lines, but their effect
on pancake vortices is less obvious.  The main questions are: a) are
pancake vortices positioned mainly inside columnar defects in the
liquid phase?, and b) are $c$-axis correlations enhanced in the
presence of CDs or do they remain similar to that in the liquid phase
of pristine crystals?

Important information about the effect of CDs on the pancake liquid
was obtained from reversible magnetization measurements
\cite{kees-Mrev}. In pristine crystals
reversible magnetization $M$ monotonically increases with the magnetic field
$B_{\bot}$. In irradiated crystals, due to gain in pinning energy,
penetration of vortices into crystals becomes more favorable than in
pristine crystals, and the diamagnetic moment drops in the presence of
CDs at low fields $B_{\bot}\ll B_{\Phi}$ when all vortices may occupy
CDs.  However, as $B_{\bot}$ increases towards the
matching field $B_{\Phi}$, interstitial vortices start to appear,
losing the advantage in pinning energy associated with the CDs.
This results in a drop of magnetization (i.e., in an increase in the
diamagnetic moment) in the field interval between $\approx B_{\Phi}/4$
and $B_{\Phi}$.  At larger fields the difference between pristine and
irradiated crystals almost vanishes.  Such an anomaly in $M(B_{\bot})$
dependence was observed in the vortex liquid up to rather high
temperature, indicating that mobile pancakes, inherent to the vortex
liquid, are localized largely onto CDs, even at temperatures close to
$T_{c}$.  Recent study of the vortex-lattice melting in weakly
irradiated Bi-2212 also confirmed this picture \cite{borya-melt}.
However, the thermodynamics of the vortex state depends weakly on
$c$-axis correlation of pancakes; it is determined mainly by intralayer
interactions, which are much stronger that those associated with the
$c$-axis vortex structure \cite{lev-magn,kosugi-bul}.  For this reason
reversible magnetization is determined by pancake concentration and
scales with $B_{\bot}$.  Thus magnetization measurements show that
pancakes are positioned mainly inside CDs at $B_{\bot}<B_{\Phi}/3$ but they
do not provide information on $c$-axis correlations of pancakes inside
CDs.

In contrast, Josephson interlayer properties of highly anisotropic
Bi-2212 are extremely sensitive to the $c$-axis correlations of the
pancakes, because Josephson current depends on the interlayer phase
difference.  Pancakes, aligned along the $c$-axis do not contribute to
the phase difference, but those, shifted due to thermal fluctuations
or pinning do.  This leads to larger phase difference in the
uncorrelated liquid and thus to the
suppression of the $c$-axis superconducting current.  Measurements of
Josephson plasma resonance, JPR, reveal for the first time that $c$-axis
correlation in the vortex liquid in the presence of CDs depends on
$B_{\bot}$ nonmonotonically in the temperature interval 60
- 69 K, showing enhancement at $B_{\bot} \sim
B_{\Phi}/3$ \cite{matsuda-JPR}, while in the liquid phase in
pristine crystals correlation drops with $B_{\bot}$ at all temperatures.
Nonmonotonic behavior with $B_{\bot}$ was found also in the dependence
of the $c$-axis resistivity, $\rho_c$, which is
also determined by interlayer
Josephson current \cite{nick-rhoc}.
Namely, $\rho_{c}$ in the same temperature
interval exhibits an increase with $B_{\bot}$ at $B_{\bot}\ll B_{\Phi}$, a
flattening or even a decrease in the field interval between
0.2~$ B_{\Phi}$ and 0.4~$ B_{\Phi}$ followed by further increase of
$\rho_{c}$ at higher $B_{\bot}$.  Unfortunately, neither JPR nor
$\rho_{c}(B_{\bot}$) measurements provide sufficient information to
estimate the enhancement of the $c$-axis correlation quantitatively.

It was shown in \cite{koshelev-JPR,nick-bx} that dependence of
 JPR frequency and $\rho_{c}$ on the parallel
component of the magnetic field provides complete information about
$c$-axis phase correlations.  Based on this idea, in this paper we study
the angular dependence of $\rho_{c}$ at different orientation of the
magnetic field $B$ in the correlated pancake
liquid phase in the presence of CDs.  We show that scaling with
$B_{\bot}$ is absent in this phase.  Moreover, we found, that at
$B=B_{cr}\approx B_{\Phi}/2$, this angular dependence is accurately
compensated, providing a crossing point where $\rho_{c}$ is
independent of the orientation $\theta$ of the magnetic field with
respect to the crystal $c$-axis over a wide range of $\theta$.  Such an
anomaly gives an opportunity to estimate the correlation length
of pancake positions along the $c$ axis at the crossing field.

For our experiments high quality Bi-2212 crystals $(T_{c}\simeq {\rm
85\:K})$ of about $1\times 1.5\times 0.02\:\ mm^{3}$ were used.  The
irradiation by 1.2 GeV ${\rm U^{238}}$-ions was performed on the
ATLAS accelerator (Argonne National Lab.).  According to TRIM
calculations these high energy ions produce in Bi-2212 crystals
continuous amorphous tracks with diameter 4-8~nm and length 25-30
$\mu$m.  Below we present the results for samples irradiated with an
effective density of columnar defects corresponding to the matching
field $B_{\Phi } = 1$~T. Another pristine sample was
used as a reference.  We checked also several crystals from a different
synthesis for universality of the obtained results.

\begin{figure}[h]																	 
\vspace{-0.3cm}																		
\hspace{-0.5cm}																		
\epsfxsize =7.7cm \epsffile{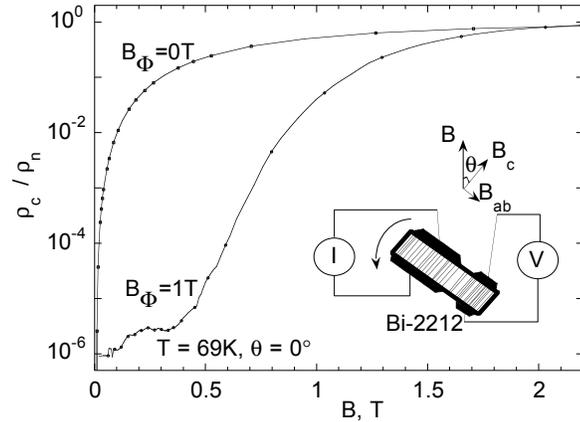}									
\caption{Dependence of c-axis magnetoresistance $\tilde{\rho_c}$ on 
c-axis field for two crystals: nonirradiated, $B_{\Phi}=0$, and 
irradiated with effective density of CDs $B_{\Phi}=1\: T$.  The latter 
displays dip due to enhanced vortex correlations in the pancake 
liquid.  Insert: schematic experimental set-up.}											   
\end{figure}

Our measurements were performed using a warm bore insert in a liquid
helium cryostat with a 9~T superconducting magnet, and also in a
liquid nitrogen, LN, dewar installed into an electromagnet that
provides a magnetic field up to 1T. The sample was attached to a
sample holder with goniometric stage, allowing rotation of the sample
with respect to the applied field with an accuracy of about $\pm
0.1^{\circ }$.  The sample was immersed in a LN bath in order to
provide accurate temperature stabilization and to avoid heating
effects. Magnetic fields applied along the {\em c}-axis of the sample
were monitored by a Hall sensor attached to the sample stage.  Two
pair of silver contacts pads were deposited on both sides of the
samples for standard 4-probe transport measurements and fine gold
leads were attached using silver epoxy.  The resistivity for the
contact pair at room temperature was about $1-3~\Omega $.
 We measured $R_c$ as a function of
applied magnetic field $B$ at different angles $\theta $ with respect
to the {\em c}-axis of the crystal (see insert
in Fig.~1).

In Fig.~1 we present the normalized magnetoresistance
$\tilde{\rho_{c}}\equiv\rho_{c}/\rho_{n}$ as a
function of applied field along the {\em c}-axis measured at $T=68$~K.
Here $\rho_n$ is the normal state resistivity at $T=120$~K.
After the sharp onset at
the irreversibility line, $\tilde{\rho_c}$ for irradiated crystals
displays a pronounced dip due to enhanced vortex correlation
associated
with filling of the CDs reported in Ref.~\onlinecite{nick-rhoc}.  At fields
$B\gg B_{\Phi}$ the magnetoresistance approaches the normal state
value.  In the pristine crystal the resistance increases with $B_{\bot}$
monotonically.

\begin{figure}[h]
\vspace{-0.5cm}																																	   
\hspace{+3cm}																																	   
\epsfxsize = 9.7cm \epsffile{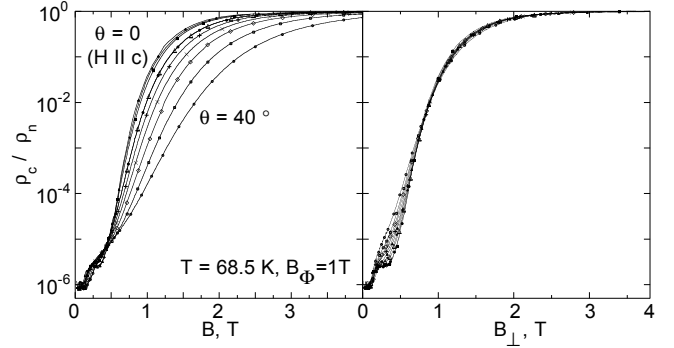}	
\vspace{-2cm}																							   
\caption{  Dependence of $\tilde{\rho_c}$ versus applied field 
$B$ for different $\theta$, progressively increased with step 
$5^{\circ}$ (left).  The same data plotted as a function of c-axis 
 field $B_c$ (right).  The scaling of $\tilde{\rho_c}$ with 
$B_{\bot}$ is broken in the correlated vortex liquid. }																																   
\end{figure}

In the left panel of Fig.~2 the family $\tilde{\rho_c}$ vs.
$B$ at different angles is presented.  The
magnetoresistance decreases with  the angle at $B>B_{\Phi}/2$ (as in
the pristine crystals), reflecting decrease of $B_{\bot}$, but {\em increases}
 with the angle at low fields $(0.1B_{\Phi}<B<B_{\Phi}/2)$,  in the
region where reentrant enhancement of $c$-axis correlation was observed
in $\rho_{c}(B_{\bot})$ dependence.

In our experiment the inclination of the sample in the applied field
results in two effects: decrease of the {\em c}-axis
field $B_{\bot}=B \,{\rm \cos} \theta $ and increase of the in-plane
field $B_{\parallel}=B \, {\rm \sin} \theta $.  In
order to separate these two effects we present $\tilde{\rho}_c$
as a function of the {\em
c}-axis component $B_{\bot}$ in the right panel of Fig.~2.
  The magnetoresistance of the sample
scales with $B_{\bot}$ at low and at high fields.  However, in the range
$B_{\Phi}/4 < B < B_{\Phi}$ this {\em scaling is broken}.  This
anomalous behavior occurs in the range of field and temperature where
enhanced vortex correlations in the pancake liquid were observed in JPR and
in $\rho_{c}(B_{\bot})$.

 In Fig.~3 we blow up the region where the
scaling is broken.  It is clearly seen on the right panel
 that $\tilde{\rho_c}$ here
depends on the in-plane field, increasing gradually with
$B_{\parallel}$.  The most outstanding feature can be seen on the left
panel of  Fig.~3.  Here $\tilde{\rho_{c}}$ curves at different angles are
presented as a function of $B$. They all cross at a single point,
which we denote as $B_{cr}$.  This crossing
point is observed for all the crystals which display the dip of
$\rho_{c}$. It means, that at $B=B_{cr}$,
{\em angular dependence of magnetoresistance practically vanishes},
 i.e., decrease
of  $\rho_{c}$ due to decrease of $B_{\bot}$ is accurately compensated
by increase of the magnetoresistance with increasing
$B_{\parallel}$.  This effect can be explained in the framework of the approach
developed in \cite{koshelev-JPR,nick-bx} and will be used below to estimate
quantitatively the characteristic length $R_{1}$ of the phase difference
in-plane correlation function.

\begin{figure}[h]	
\vspace{-0.5cm}												
\hspace{1cm}																
\epsfxsize = 9.7cm \epsffile{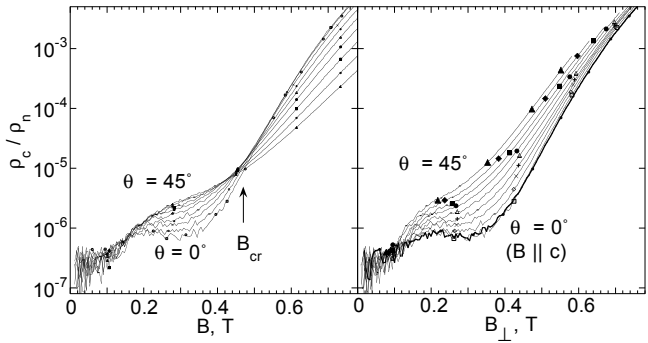}	
\vspace{-2cm}							 
\caption{Blow up of the anomaly for the irradiated crystal.
 Dependence of 
$\tilde{\rho_c}$ versus applied field $B$ (left) and versus c-axis field 
$B_{\bot}$ (right).  $\theta$ increases from 0 to $45^{\circ}$ with step 
$5^{\circ}$. The crossing point $B_{cr}$ corresponding to the absence of 
the angular dependence is marked on the $\tilde{\rho_c}$~vs.~$B$ 
panel.}																	 
\end{figure}			

Let us first discuss how the in-plane field component affects the {\em
c}-axis magnetoresistance.  An approach developed initially for the
field behavior of JPR frequency \cite{koshelev-JPR} was extended recently
for $c$-axis transport properties \cite{nick-bx}.  In Josephson coupled
superconductors in the presence of a $c$-axis current, the voltage
$V_{n,n+1}$ is
induced by slips of the phase difference $\varphi_{n,n+1}({\bf r},t)$
between the layers $n$ and $n+1$, as described by the Josephson relation
$V_{n,n+1}=(\hbar/2e)\dot{\varphi}_{n,n+1}$.  Here
${\bf r}=x,y$ are coordinates in the $ab$ plane, and $t$ denotes the
time.  The $c$-axis conductivity in the vortex liquid state
$\sigma_{c}=1/\rho_{c}$ is determined by the Kubo formula
\cite{koshelev-kubo}
\begin{eqnarray}
\sigma_c(B_{\bot},B_{\parallel})=(s{\cal J}_0^2/2T)\int_0^{\infty}dt
\int d{\bf r}S({\bf r},t),
\label{sig} \\
S({\bf r},t)=2\langle\sin\varphi_{n,n+1}(0,0)
\sin\varphi_{n,n+1}({\bf r},t)\rangle
\nonumber \\
\approx\langle\cos[\varphi_{n,n+1}({\bf r},t)-\varphi_{n,n+1}(0,0)]\rangle,
\label{1}
\end{eqnarray}
where ${\cal J}_0$ is the Josephson critical current, $s$ is the
interlayer distance, and $<\ldots>$ means thermal average and average
over disorder.  Time variations of the phase difference are
caused mainly by mobile pancakes \cite{koshelev-SC} induced by $B_{\bot}$,
while the parallel field component induces a stationary phase difference in
 the lowest order in Josephson coupling. We split
$[\varphi_{n,n+1}({\bf r},t)-\varphi_{n,n+1}(0,0)]$ into the
contribution induced by pancakes and that caused by the unscreened
parallel component $B_{\parallel}$.  Assuming that $B_{\parallel}$ is
along the $x$ axis, we obtain:
\begin{eqnarray}
&&\varphi_{n,n+1}(0,0)-\varphi_{n,n+1}({\bf r},t)= \label{me} \\
&&[\varphi_{n,n+1}(0,0)-\varphi_{n,n+1}({\bf r},t)]_{B_{\parallel}=0}-
2\pi sB_{\parallel}y/\Phi_0.
\nonumber
\end{eqnarray}
In a single point-like junction the phase difference induced by
$B_{\parallel}$ results in the Fraunhofer pattern of Josephson
critical current as a function of the magnetic
field parallel to the junction.  In our case of a multilayer
superconductor the phase difference induced by $B_{\parallel}$
interferes with that induced by pancakes. Thus we obtain
\begin{eqnarray}
&&\sigma_c(B_{\bot},B_{\parallel})=(\pi s{\cal J}_0^2/T)\int
drr\tilde{G}(r,B_{\bot})
J_0(\alpha B_{\parallel}r),
\label{me1} \\
&&\tilde{G}(r,B_{\bot})=\int_0^{\infty}dtS(r,t,B_{\bot}), \nonumber
\end{eqnarray}
where $J_0(x)$ is the Bessel function, $\alpha=2\pi s/\Phi_0$
($\Phi_{0}$ is a flux quantum), and the
function $\tilde{G}({\bf r},B_{\bot})$ depends on correlations
of the phase difference induced by pancake vortices.
For small $B_{\parallel}$ (small angles) we expand
the Bessel function in $B_{\parallel}$:
\begin{eqnarray}
&&\sigma_{c}(B_{\bot},B_{\parallel}) \approx
\sigma_{c}(B_{\bot},0)\left[ 1-\frac{1}{4}\alpha^{2}{R_{1}^{2}(B_{\bot})}B_{\parallel}^2
\right], \label{rel}
\\
&&R_1^2(B_{\bot})=\int drr^3\tilde{G}(r,B_{\bot})/\int drr\tilde{G}(r,B_{\bot}).
\nonumber
\end{eqnarray}
Here the in-plane correlation length
$R_{1}(B_{\bot})$ describes the decay of the phase difference correlation
function. When pancakes are positioned mainly inside CDs, the characteristic
length of this decay, $R_1$, gives direct information on the $c$-axis
correlation of pancake positions because drop of the phase
difference correlations in the $ab$ plane
is caused by interruptions in the pancake arrangement
along CDs \cite{koshelev-JPR}. The characteristic length, $L$, of the pancake
density correlation function is related to $R_1$ as
$L/s\approx R_1^2/10a^2$, where $a=(\Phi_0/B_{\bot})^{1/2}$ is the
intervortex distance \cite{nick-bx}. This expression and Eq.~(\ref{rel})
are a key points for further  discussion.

It is clear from Eq.~(\ref{rel}) that generally $\sigma_{c}$ depends
upon both components of the field.  However, in an uncorrelated
liquid, when $R_{1} \approx a$,
the effect of the in-plane field is small
and can be observed only in high fields $B_{\parallel}\gtrsim\Phi_{0}/sa$.
As a
result, in the uncorrelated pancake liquid the {\em c}-axis
conductivity scales with $B_{\bot}$ in fields $B\ll\Phi_{0}/sa$.  For an
irradiated sample, as filling of CDs progresses, vortices start to
form stacks and the correlation length $R_{1}$ significantly exceeds
$a$, reaching a maximum value near $B_{\Phi}/3$.  Here the
effect of the in-plane field becomes significant and scaling of
magnetoresistance with $B_{\bot}$ breaks.  As the field $B_{\bot}$ further
increases, the fraction of the interstitial vortices increases.
These vortices introduce additional disorder to the system which
results in decay of correlations, in reduction of $R_{1}$ down to
$\sim a$
and, consequently, the effect of $B_{\parallel}$ drops.  Then, at
elevated fields above $B_{\Phi}$ scaling of $\rho_{c}$ with $B_{\bot}$ is
restored.  This scenario describes qualitatively well our experimental
results.

Now we show that the crossing point in Fig~3 allows us to estimate
the correlation radius $R_{1}(B_{cr})$. Using Eq.~(\ref{rel})
for the variation of $\sigma_{c}$ at small angles we expand as:
\begin{equation}
\delta \sigma _{c}(B_{\parallel},B_{\bot}) \approx
\frac{\partial^2\sigma_{c}}{\partial B^{2}_{\parallel}}\:\frac{(\delta
B_{\parallel})^2}{2}+
\frac{\partial\sigma_c}{\partial B_{\bot}} \delta B_{\bot}.
\end{equation}
Substituting $B_{\parallel} \approx B\theta$ and
$B_{\bot} \approx B(1-\theta^2/2)$ we obtain
\begin{equation}
\delta \sigma _{c}(B_{\parallel},B_{\bot}) \approx \frac{1}{2}
B\left(\frac{\partial^2\sigma_{c}}{\partial B^{2}_{\parallel}}\:B-
\frac{\partial\sigma_c}{\partial B_{\bot}}\right) (\delta\theta)^2.
\label{cross}
\end{equation}

Independence of $\sigma_{c}$ on the angle $\theta$ at $B_{cr}$ occurs
when the expression in the brackets becomes zero.
With the help of Eq.~(\ref{rel}) we obtain the correlation radius $R_{1}$:
\begin{equation}
R_1^2(B_{cr})=-\frac{1}{2B_{cr}}\: \frac{\Phi_0^2}{\pi^2s^2 } \:
	\left[\frac{\partial \ln \sigma_c (B_{\bot},0)}{\partial B_{\bot}}\right]_{B_{\bot}
=B_{cr}}.
\label{radius}
\end{equation}
From the
data presented in  Fig.~3, we obtain $(\ln \sigma_c (B_{\bot},0)
)^\prime \approx -23 \: {\rm T^{-1}}$ at $B=B_{cr} \approx 0.47$~T.
Using Eq.~(\ref{radius}) we calculate $R_{1}/a \approx 32$.  This value is
in reasonable agreement with the result obtained in \cite{nick-bx},
where $R_{1}\sim 10a$ was found. Thus the correlations length $L$ of pancakes
positions along the $c$ axis is $\approx 100s$.

We would like to emphasize the difference between our observation and
results obtained in flux-transformer geometry
\cite{richard}.  In those latter experiments current was applied
to the top surface of the sample and voltage was measured in the both
top and bottom layers.  In some range of the magnetic fields and
temperatures top and bottom voltages coincide, indicating a similar
motion of vortices in all layers.  This was considered as evidence of
pancake coupling along the $c$-axis.  Unfortunately, the situation
here is rather complicated because of mixing of the in-plane and
$c$-axis resistivities.  Namely, the current applied along the top layer
penetrates deep into the sample along the $c$ axis and flows along the
bottom layer as well as along the top layer.  This results in a
similar motion of the pancakes through the full sample thickness, even
without $c$-axis correlation provided $\rho_{c}$ is small enough.  As we have
shown, the $c$-axis transport measurements as a function of in-plane
field component are free of this additional effect and are very sensitive
to pancake correlation between adjacent layers.

To conclude, we have presented evidence for the presence of a
partially aligned vortex liquid in irradiated Bi-2212 from the angular
dependence of {\em c}-axis magnetoresistance.  In the range of
temperatures and magnetic fields where $c$-axis correlations develop,
interlayer transport becomes much more sensitive to the in-plane
component of the magnetic field in comparison with the uncorrelated
liquid in pristine crystals.  Then scaling of $\rho_c$ with $B_{\bot}$
breaks, and, at the field $B_{cr}$, angular dependence becomes very
weak.  From this crossing point we estimate the $c$-axis correlation length
of pancake positions $L/s\approx 100$ at $B_{\bot}\approx B_{\Phi}/2$.
The origin of this alignment can be due to magnetic intralayer
interaction of pancakes which favors similar filling of CDs due to
their geometry.  However an additional effect of interlayer magnetic
interaction of pancakes cannot be excluded so far.

Fruitful discussions with H.~Safar, and Y.~Matsuda are greatly
acknowledged.  We thank J.~H.~Cho for providing the Bi-2212 single
crystals.  We also thank K.E. Gray, J.U. Lee, D.H. Kim, and D.J.
Hofman of Argonne National Lab. for the help in irradiation of the samples.
This work is supported by U.S. DOE.

\end{document}